
\input harvmac

\def\ymh{Yang-Mills-Higgs}
\def\ie{{\it i.e.}}

\def\ib{{\bar I}}
\def\xb{{\bar x}}
\def\ab{{\bar a}}
\def\nb{{\bar N}}
\def\co{{\cal O}}
\def\cd{{\cal D}}
\def\cc{{\cal C}}
\def\xp{{x^\prime}}
\def\xpb{{\xb^\prime}}
\def\calh{{\cal H}}
\def\calr{{\cal R}}
\def\ay{A_{\rm Yung}}

\def\iib{instanton-antiinstanton pair}

\Title{\vbox{\hbox{UTTG-18-94}}}
     {\centerline{On Multiple Instanton-antiinstanton Configurations}}

\centerline{Meng-yuan Wang\footnote{$^\dagger$}{meng@physics.utexas.edu}}
\bigskip\centerline{Theory Group, Department of Physics,
University of Texas, Austin, TX 78712}

\vskip .3in
We show why and how the $I\ib$ valley trajectory commonly used in the
literature so
far is in fact unsatisfactory. A better $I\ib$ valley is suggested.
We also give analytic expressions for the multiple instanton-antiinstanton
configurations in the pure Yang-Mills theory.
These formulas make it possible to go beyond the
dilute gas approximation and calculate
the multi-body interactions among instantons and antiinstantons.

\Date{12/94}

\newsec{Introduction}
In the Euclidean version of the 4-dimensional pure
Yang-Mills theory, with the action
\eqn\ymaction{S\ =\ {1\over{2g^2}}\int d^4x{tr}F^2,}
field configurations which correspond to finite values of the action
fall into discrete sectors characterized by an integer $Q$, the
Pontryagin index. In each sector,
the solution to the field equation is exactly known
and shown to be unique.
They are the $N$-instanton ($I^N$)
solutions\nref\atiyah{M.F.Atiyah, {\it The
Geometry of Yang-Mills Fields}.}\nref\atiyaha{M.F.Atiyah, V.G.Drinfeld,
N.J.Hitchin and Y.I.Manin, {\sl Phys. Lett.} {\bf 65A}
(1978) 185.}\nref\drinfeld{V.G.Drinfeld and Y.I.Manin,
{\sl Funk. Analiz.} {\bf 13}
(1979), 59.}\refs{\atiyah{--}\drinfeld}\foot{We will
adopt the quaternion notation used in ref.\xref\atiyah. For a brief
introduction, see Appendix $A$.}, with $N=Q$.
The $I^N$ solution is parametrized by $(8N-3)$ independent degrees of
freedom, which we shall denote as $\omega$. One can
interpret them as the positions (4 for each instanton), the sizes (1 each) and
the group orientations (or phases, 3 each,
minus the 3 overall phases which can be undone by
the global gauge transformations). If we are interested in the $Q$-sector
contribution to the path integral,
\eqn\qpi{Z_Q\ =\ \int_{A\in Q{\rm sector}}[\cd A]\exp(-S[A]),}
the $I^N$ will dominate because it minimizes the action.
Furthermore, those $(8n-3)$ zero modes should be isolated from the
other degrees of freedom
using the collective coordinate method, together with the 3 global gauge
transformations. Keeping only the supposedly dominant exponent\foot{This
is a very crude approximation. It gives only part of the leading contribution
in the semi-classical expansion. This is adequate for our purpose, however.
A general treatment for the complete leading term of
the semi-classical approximation can be found in our
other paper\ref\wang{M.Wang,
UT preprint UTTG-21-94}. The explicit calculation for the one-instanton case
was first carried out by 't Hooft\ref\thooft{G.'t Hooft,
{\sl Phys. Rev.} {\bf D14}
(1976) 3432.} and can be found in numerous reviews.}, we have
\eqn\qpia{Z_Q\ \sim\ {1\over{Q!}}\int d^{8Q}\omega\ \ \exp(-S(\omega)).}
The original field theory problem is thus reduced to that of interacting
particles. In this specific case, the action $S$ is well known, \ie
\eqn\inaction{S_{(I^N)}\ =\ N S_I\ =\ {{8\pi^2}\over{g^2}}N,}
independent of $\omega$.
In other words, it consists purely of instanton
``self-action", and there is no interaction among instantons.
As for the path integral, it now becomes
\eqn\qpib{Z_Q\ \sim\ {1\over{Q!}}\left\{\int d^{8}\omega\exp(-S_I)\right\}^Q.}

We have ignored those sub-dominant configurations which are not solutions
to the field equation. Their contributions may be important sometimes and
should be included in our approximation. The most important of these
sub-dominant configurations is the
$N$-instanton-$\nb$-antiinstanton ($I^N\ib^\nb$), with
$N-\nb=Q$. For widely separated $I^N\ib^\nb$, the
interactions are negligible and
we again have
\eqn\inibnbaction{S_{(I^N\ib^\nb)}\ \sim\ (N+\nb) S_I.}
Therefore,
\eqn\qpib{Z_Q\ \sim\ \sum_{N,\nb}{{\delta_{N-\nb-Q}}\over{N!\nb !}}
\int d^{8(N+\nb)}\omega\ \ \exp(-S(\omega)).}
Using the identity
\eqn\ida{\delta_n\ =\ \int_0^{2\pi}{{d\theta}\over{2\pi}}e^{-in\theta},}
we can further simplify \qpib,
\eqn\qpic{\eqalign{Z_Q\ &\sim\ \int_0^{2\pi}d\theta{{e^{iQ\theta}}\over{2\pi}}
\sum_{N,\nb}{1\over{N!\nb !}}
\int d^{8(N+\nb)}\omega\ \ \exp(-N(S_I+i\theta)-\nb(S_I-i\theta))\cr
&\sim\ \int_0^{2\pi}d\theta{{e^{iQ\theta}}\over{2\pi}}
\exp\left(\int d^8\omega\ e^{-S_I}\ \cos\theta\right).\cr}}

We have presented a simplified version of the so-called dilute-instanton-gas
calculation.
There are three possible improvements over \qpic.
Firstly, one can incorporate the
pre-exponential factor so that the result becomes the true leading term in the
semi-classical expansion. This has been done for $Q=1$ in ref.\xref\thooft.
Secondly, one may want to improve \inibnbaction\ by introducing
instanton-antiinstanton interactions. This will be the main goal of this
paper. Lastly, since the integral $\int d^8\omega$ contains the integration
over the instanton size $\rho$, the semi-classical (small $g$)
approximation naturally
breaks down at the infrared limit, as a result of the
renormalization group running
effect. This is a common problem that plagues all
semi-classical treatments for
the 4-dimensional pure Yang-Mills theory. We are
unable to provide new insights
into this problem. However, this difficulty is a
totally separate issue from the
dilute-gas approximation we will
try to improve, and should not invalidate our
treatments. If the theory contains a
scale cutoff as a result of the Higgs mechanism
or finite temperatures, the infrared problem is circumvented and our multiple
instanton-antiinstanton results will be valid. For simplicity, we will avoid
for now the complexity involved in the Yang-Mills-Higgs system, which will be
discussed in our future paper\wang,
and concentrate on the pure Yang-Mills theory
instead.

There have been some previous efforts
trying to go beyond the dilute-instanton-gas
approximation. Callan, Dashen and
Gross\ref\callan{C.G.Callan, R.Dashen and D.J.Gross,
{\sl Phys. Rev.} {\bf D17}
(1978) 2717;\ {\bf D19} (1979) 1826.} were the first to
compute the leading $I\ib$ interaction at the large separation ($R$) limit.
Their result, however, is
not conformally invariant. It is also very difficult to
calculate subleading terms using their method. Superior in both aspects is the
later work by Yung\ref\yung{A.Yung, {\sl Nucl. Phys.} {\bf B297} (1988) 47.}.
Using a spherical ansatz, he reduced the Yang-Mills action to that of a
quantum mechanical double-well. This trick enabled him to write
down the $I\ib$ configuration in the Yang-Mills theory corresponding to the
kink-antikink in the double-well system, and the $I\ib$ interaction to
all orders in $\rho/R$ simply followed. We will review this important result
in detail in Section 2.

Elegant though Yung's solution is, it relies heavily on the coincidence which
connects the Yang-Mills theory with the simpler quantum mechanical system,
which in turn relies on
the spherical ansatz. Therefore this method obviously cannot
be generalized to anything more complex than $I\ib$. Employing a brand new
philosophy, we construct a systematic treatment that will make it possible
to find expressions for $I^N\ib^\nb$. In Section 3, we illustrate this method
in the simplest case of $I\ib$. Surprisingly, Yung's solution will be shown
to be unsatisfactory. This is an important result because naive application of
Yung's valley formula has
been heavily used to compute the high-energy baryon-number
violating cross-section in
the standard model. Improvement on the understanding of
the valley trajectory can dispel some common misconceptions.
In Section 4, we generalize our result to $I^2\ib^2$ and
beyond. Although these semi-classical results do not have direct applications
in QCD at this moment (except maybe for the instanton-liquid hypothesis), they
are nonetheless interesting not only because they
provide corrections to the dilute-instanton-gas approximation, but also
because they can serve as a
primer for similar treatments for the Yang-Mills-Higgs
system. It has been argued by some authors\nref\zakharov{V.Zakharov,
{\sl Nucl. Phys.} {\bf B353} (1991) 683; {\sl Nucl. Phys.}
{\bf B377} (1992) 501.}\nref\maggiore{M.Maggiore and M.Shifman,
{\sl Phys. Rev.} {\bf D46} (1992) 3550; {\sl Nucl. Phys.}
{\bf B371} (1992) 177.}\refs{\zakharov,\maggiore} that, again,
in the high-energy instanton-induced baryon-number
violating processes, the multi-instanton
effects become important long before the
one-instanton amplitude has grown large. Their analysis relied, however, on a
crude nearest neighbor approximation, and was questioned by
other authors\nref\dorey{N.Dorey and
M.Mattis, {\sl Phys. Lett.} {\bf B277} (1992)
337.}\nref\kripfganz{V.V.Khoze, J.Kripfganz and A.Ringwald, {\sl Phys. Lett.}
{\bf B275} (1992) 381; {\sl Phys. Lett.} {\bf B277} (1992)
496.}\refs{\dorey,\kripfganz}.
This controversy clearly cannot be settled until we gain better knowledge
of the multi-instanton configurations in the Yang-Mills-Higgs theory, and
our result should be the first step toward achieving this goal.

\newsec{Yung's Valley Solution for $I\ib$}
Before we introduce Yung's result, it is
necessary to familiarize ourselves with
the conformal properties of the instantons.
This is because the 4-dimensional
Yang-Mills lagrangian is classically
invariant under the conformal group, which
includes the Poincare group as well as the dilatation and four special
conformal transformations. Together with the global gauge transformations,
they ensure that all 8 parameters of
the one-instanton solution ($I$) correspond to
zero modes. One can apply this group theory analysis to the two-instanton
solution\foot{The standard introductory text for the the $I^N$
solution is by Atiyah\atiyah. We
discuss some interesting properties of $I^2$ in
one of our earlier papers\ref\wanga{M.Wang,
UT preprint UTTG-17-94}.} ($I^2$) also,
and find that, of the 16 parameters, all
but two have to be zero modes due to these
symmetries. Although one can show that
even these two potential exceptions turn out
to be zero modes, either by direct
computation or by using a much more involved
argument than we care to reproduce here, it is still interesting
to find these two modes explicitly. After some
tedious calculation, we find that they correspond to the relative phase
and a dimensionless parameter
$z_2=(R^2+\rho_1^2+\rho_2^2)^3/(R^2\rho_1^2\rho_2^2)$
which can be interpreted as the separation between the two instantons.

Since the $I\ib$ configuration
should also be described similarly by 16 parameters,
it is natural to wonder what insight
we can get using the group theory argument.
This turns out to be more difficult than one would imagine because of the
uncertainty involved in reducing a
field configuration with an infinite number
of degrees of freedom, to an unknown
expression parameterized by only 16. We therefore
make the assumption that $I$ and $\ib$
can be put together in a more or less
linear manner\foot{We are aware that this sounds awfully vague. We do not
consider it worthwhile to present this result in detail though, because
its importance has been largely diminished by the results we shall present
later in this paper.}, and find again that all but two correspond to zero
modes. These two possible non-zero modes are the relative phase and
the parameter,
\eqn\defineg{z\ =\ (R^2+\rho_1^2+\rho_2^2)/(2\rho_1\rho_2).}
They are invariant under all the conformal transformations.
Let us ignore the relative phase for
now, and concentrate on the $z$ direction.
As $z$ increases from its minimal value
1 to infinity, we produce an \iib\ from
the trivial vacuum and pull them farther and farther apart. Therefore, the
action should increase from 0 to $2S_I$
accordingly. This is exactly what makes
the \iib\ important. The action flattens
out at large separation, and its effects
would have been badly accounted for if we
had naively treated this mode like
any other quantum perturbation. Yung and other authors call this mode the
valley direction because it corresponds
to a low-lying valley if one considers
the action as a functional in the field configuration space. We will also
use the phrase ``quasi-zero modes" sometimes, partly because they require
the collective coordinate treatment, similar to the real zero modes.

We are now ready to present Yung's result. Making full use of the conformal
symmetries, we can tranform any given set of $I\ib$ parameters into
one which satisfies
\eqn\yungslice{R=0,\ \ \ \ \ \ \rho_1\rho_2=1,\ \ \ \ \ \ \rho_1\le\rho_2.}
This corresponds to an instanton sitting
right on top of an antiinstanton of a
possibly different size. Therefore, it is natural to make the following
spherical ansatz,
\eqn\yansatz{A_{\rm Yung}\ =\ {\rm Im}\left\{{{xd{\bar x}}\over
{x^2}}s(x^2)\right\},}
since both the instanton and the antiinstanton can be put in this form.
More specifically, the instanton has to be put in the regular gauge,
\eqn\rgauge{A_I^{\rm reg}\ =\ {\rm Im}\left\{{{xd{\bar x}}\over
{x^2+{1\over{\rho_2^2}}}}\right\},}
and the antiinstanton in the singular gauge,
\eqn\sgauge{A_\ib^{\rm sing}\ =\ {\rm Im}\left\{{{\rho_2^2xd{\bar x}}\over
{x^2(x^2+\rho_2^2)}}\right\}.}

Our next step is to substitute \yansatz\
into \ymaction. Here a miracle occurs,
and we find
\eqn\dwaction{S\ =\ {{48\pi^2}\over{g^2}}\int_{-\infty}^\infty
dt\left\{{1\over2}\left({{ds}\over{dt}}\right)^2+{1\over2}
\left[\left(s-{1\over2}\right)^2-{1\over4}\right]^2\right\},}
where $t=\ln{x^2}$. As promised earlier, the integral is exactly the
action of a quantum mechanical double-well. The instanton \rgauge\
gives the kink at $-\xi$,
\eqn\kink{s_I^\xi(t)\ =\ {1\over2}\left[1
+\tanh\left({1\over2}(t+\xi)\right)\right],}
where $\xi=\ln{\rho_2^2}$, and the antiinstanton \sgauge\ gives
the antikink at $\xi$,
\eqn\antikink{s_\ib^\xi(t)\ =\ {1\over2}\left[1-\tanh\left({1\over2}(t-\xi)
\right)\right].}
Such one-to-one
correspondences are encouraging, and one is naturally tempted to
use the kink-antikink configuration for $I\ib$. We then have
\eqn\kak{\eqalign{s\ &=\ {1\over2}\left[\tanh\left({1\over2}(t+\xi)\right)
-\tanh\left({1\over2}(t-\xi)\right)\right]\cr
&=\ {{x^2}\over{x^2+\rho_1^2}}-{{x^2}\over{x^2+\rho_2^2}}.\cr}}
Putting this back into \yansatz, we have
\eqnn\rmr
\eqnn\sms
$$\eqalignno{&A_{\rm Yung}^{r-r}\ =\ {\rm Im}\left\{
{{xd{\bar x}}\over{x^2+\rho_1^2}}-{{xd{\bar x}}\over
{x^2+\rho_2^2}}\right\},&\rmr\cr
=\ &A_{\rm Yung}^{s-s}\ =\ {\rm Im}\left\{-
{{\rho_1^2xd{\bar x}}\over{x^2(x^2+\rho_1^2)}}
+{{\rho_2^2xd{\bar x}}\over{x^2(x^2+\rho_2^2)}}\right\},&\sms\cr}$$
or, after a gauge transformation,
\eqn\spr{A_{\rm Yung}^{s+r}\ =\ {\rm Im}\left\{
{{\rho_1^2{\bar x}dx}\over{x^2(x^2+\rho_1^2)}}+
{{{\bar x}dx}\over{x^2+\rho_2^2}}\right\}.}
Notice that since $z=(\rho_1^2+\rho_2^2)/2$ and
$\rho_1\rho_2=1$, $\ay$ describes
the trivial vacuum for $z=1$ (as can be seen from \rmr), and an \iib\ at
large separation for $z\rightarrow\infty$ (from \spr). This is just what
one would expect from the $I\ib$ valley. Substituting \kak\ into \dwaction,
we get the action profile for Yung's $I\ib$ valley,
\eqn\iibaction{S(\ay)\ =\ {{16\pi^2}\over{g^2}}\left\{
{{\rho_2^8-8\rho_2^4-17}\over{(1-\rho_2^4)^2}}-{{36\rho_2^4+12}
\over{(1-\rho_2^4)^3}}\ln\rho_2^2\right\}.}

As explained earlier, $\ay$ is given only for
the \iib s satisfying \yungslice.
The expression for a general \iib\ with
arbitrary $(R^0,\rho_1^0,\rho_2^0)$ is found by
conformal-transforming the corresponding $\ay$ with $z=(\rho_1^2+\rho_2^2)/2=
({R^0}^2+{\rho_1^0}^2+{\rho_2^0}^2)/(2\rho_1^0\rho_2^0)$.
The action for a general \iib\ is therefore
identical to that of the corresponding
$\ay$, which can be expressed in terms of $z$ by substituting
\eqn\zrho{\rho_2^2\ =\ z+\sqrt{z^2-1},}
into \iibaction. We have
\eqn\iibactiona{\eqalign{S_{\rm Yung}(z)\ =\ &{{16\pi^2}\over{g^2}}\left\{
{{2-8z^2+9z\sqrt{z^2-1}}\over{z^2-1}}\right.\cr
&\left.+{{3\left(2z^3-(2z^2+1)\sqrt{z^2-1}\right)}
\over{(z^2-1)^{3\over2}}}\ln\left(z+\sqrt{z^2-1}\right)\right\}.\cr}}

If this derivation for analytic $I\ib$ expressions seems amazingly simple, it
is because we have not mentioned the caveat yet. As is well known, the
$I\ib$ valley, or any quasi-zero mode in general, is not a minimum of the
action, or equivalently a solution to the field equation,
\eqn\feq{\left.{{\delta S(A)}\over{\delta A}}\right|_{A_{I\ib}}\ =\ 0.}
Instead, it is the minimum only under constraints which limit the
degree of freedom along the valley direction. Therefore, the valley
configuration $A_{I\ib}$ is a solution to \feq\ under a certain constraint.
Yung considered the following constraint to be natural,
\eqn\yc{\int d^4x \left(A-A_{I\ib}\right)
{{\partial A_{I\ib}}\over{\partial z}}\ =\ 0,}
because the sectors in which the solution $A_{I\ib}$ is a minimum are
perpendicular to the valley direction.
One therefore has to solve
\eqn\yca{\left.{{\delta S(A)}\over{\delta A}}\right|_{A_{I\ib}}\ \propto\
{{\partial A_{I\ib}}\over{\partial z}}.}
Unfortunately, the Yung form \rmr\ or \spr\ does not satisfy \yca.
One is thus forced to consider constraints which cut out sectors not
perpendicular to the valley direction, or
putting it differently, perpendicular
only if one defines a generalized inner product which varies with
$z$. This is why Yung correctly limited the validity of his result
to the leading order result in the large $z$ region only.
Other authors have been more daring\ref\khoze{V.V.Khoze
and A.Ringwald, CERN preprint, CERN-TH-6082-91.}. They
claim that with a suitably
defined varying inner product, $\ay$ should be considered
a valid valley trajectory
for all values of $z$.
This turns out not to be true, as we shall see in the next section.

\newsec{The Valley Method Done Right}
Although the correspondence between the Yang-Mills
instantons and the kinks in
the double-well potential is an amazing fact, it also prevents us from
generalizing
Yung's method to anything not spherically symmetric. In order to overcome
this difficulty, we have to find a way to deal with
the Yang-Mills instantons
directly. Let us reexamine Yung's derivation for inspiration.
Notice that the kink-antikink configuration we used in \kak\
does not satisfy the analog of \yca\ in the double-well system. Instead,
it is simply a linear combination of a kink and an antikink. In fact, this
is why $\ay$ does not satisfy \yca\ and requires a redefinition of the
inner product. One may wonder if Yang-Mills instantons and antiinstantons
can be put together linearly
without us bothering with their quantum mechanical
counterparts.

Such attempts have been made since the early days of instantons.
They inevitably failed because as the \iib\ gets close to each other,
the expression will not gradually approach the trivial vacuum, if one
insists on having both in the same gauge, which most of the early authors
did. If we reason carefully, however, we find no real reason why
this has to be so, other than the fact that it would automatically guarantee
the $Z_2$ spacial reflection symmetry of the lagrangian. We will
abandon this reflection symmetry in order to pursue a simple expression for
the valley configuration. This expression must satisfy all other good
properties one would expect from the \iib. We now list these criteria,
\vskip 0.2in
\item{1)} $A_{I\ib}$ belongs in the $Q=0$ sector.
\item{2)} $A_{I\ib}$ has easily identifiable instanton
parameters, and covers the entire 16-dimensional
parameter space spanned by all zero- and nonzero-modes.
\item{3)} $A_{I\ib}$ becomes the sum of an instanton and an
antiinstanton at large separation, and approaches the trivial vacuum
as $z\rightarrow1$.
\item{4)} $A_{I\ib}$ respects the symmetries
of the theory. This includes
the conformal symmetries and a $Z_2$ symmetry which
we will explain in more detail later.
\vskip 0.2in
These criteria may seem arbitrary, but in fact they are not. They are all
that we know for sure about $A_{I\ib}$. Every other detail in $A_{I\ib}$
can be compensated by the choice of constraints.
To see this, recall that $A_{I\ib}$ satisfies \feq\ only after a contraint is
applied. If we choose a general linear constraint
\eqn\lc{\int d^4x \left(A-A_{I\ib}\right)f_z(x)\ =\ 0,}
what we need to solve becomes
\eqn\lca{\left.{{\delta S(A)}\over{\delta A}}
\right|_{A_{I\ib}}\ \propto\ f_z(x).}
Instead of fixing the constraint to solve for $A_{I\ib}$, which is always
a difficult if not impossible task, we can choose $A_{I\ib}$ first, then use
\lca\ to find $f_z$, which amounts to no more than a simple substitution
of $A_{I\ib}$ into the left hand side of \lca. This is to say that
the bottom of the valley is not strictly-defined, and we
should make the best use of this freedom.

Before we endeaver to find the expression satisfying all these criteria, let's
first examine how $\ay^{s+r}$ stacks up against them. It satisfies $Cri.1$ and
$Cri.2$ quite trivially, although we haven't mentioned how to put in
the phases. This is done by sandwiching both the instanton and the
antiinstanton with $SU(2)$ group elements, or in our notation, unit quaternion
constants $a$ and $b$, as follows,
\eqn\spra{A_{\rm Yung}^{s+r}\ =\ {\rm Im}\left\{
{{\rho_1^2a{\bar x}dx{\bar a}}\over{x^2(x^2+\rho_1^2)}}+
{{b{\bar x}dx{\bar b}}\over{x^2+\rho_2^2}}\right\}.}

As for $Cri.3$, $\ay^{s+r}$ satisfies the first part because it is simply a
linear combination of the (anti)instantons, and the second part because
the (anti)instantons are in the singular and the regular gauge respectively.
When $z\rightarrow1$, \spr\ becomes
\eqn\pg{\eqalign{A_{\rm Yung}^{s+r}\ &=\ {\rm Im}\left\{
{{{\bar x}dx}\over{x^2(x^2+1)}}+{{{\bar x}dx}\over{x^2+1}}\right\}\cr
&=\ {\rm Im}\left\{{{{\bar x}dx}\over{x^2}}\right\},\cr}}
which is a pure-gauge configuration.\foot{In fact, this coincidence is
more general than this, as we shall see in the next section.}

So far, $\ay^{s+r}$ has passed the tests with flying colors. This suggests
that it is pretty close to the ``true" valley
bottom. Unfortunately, as we shall
show now, it is not close enough. The problem lies in $Cri.4$. $\ay^{s+r}$
does respect the conformal symmetries, but this is done in a rather
artificial way. Recall that $\ay^{s+r}$ is defined only under the constraint
\yungslice. All other configurations are given by conformal projection.
Although this seems contrived, it nonetheless gets the job done.
It is not so when it comes to the $Z_2$ symmetry, by which we mean exchanging
$\rho_1$ and $\rho_2$. Clearly this corresponds to exchanging the instanton
and the antiinstanton, and thus the action should remain unchanged\foot{This
is a weaker form of the spacial reflection symmetry. Instead of the lagrangian,
we only require the action to be invariant.}. It turns out that $S_{\rm Yung}$
does not respect this symmetry\foot{In fact,
it is possible to have $S_{\rm Yung}$
compatible with the $Z_2$, but this is done by defining $A_{\rm Yung}$ as in
Eq.\rmr,\sms\ and \spr\ only
for $\rho_1\le\rho_2$. One
then defines the configurations with
$\rho_1\ge\rho_2$ to be the $Z_2$ projections
of $A_{\rm Yung}$. Unfortunately, this
procedure introduces a discontinuity into
$S_{\rm Yung}$ at $z=1$.}. The problem is particularly bad
for $z\sim1$. Let's first define
\eqn\dtheta{\theta=\rho_2-\rho_1.}
Expanding \iibaction\ for small $\theta$ then gives
\eqn\iibactionb{S_{\rm Yung}\ \sim\ {{16\pi^2}\over{g^2}}\left\{
{6\over5}\theta^2-{4\over5}\theta^3+{9\over{35}}\theta^4+\co\left(
\theta^5\right)\right\}.}
The odd power terms clearly violate the $Z_2$ symmetry. If problems
in the third power don't seem too bad, consider the action
$S_{\rm Yung}$ for the \iib\ with
opposite phases,\ie\ $a{\bar b}+b{\bar a}=0$.
We have
\eqn\iibactionc{S_{\rm Yung}^{+-}\ =\ {{16\pi^2}\over{g^2}}\left\{
{{\rho_2^4+1}\over{\rho_2^4-1}}-{{4}
\over{(1-\rho_2^4)^2}}\ln\rho_2^2\right\}.}
This has the small $\theta$ expansion,
\eqn\iibactiond{S_{\rm Yung}^{+-}\ \sim\ {{16\pi^2}\over{g^2}}\left\{
2-{2\over3}\theta+{1\over6}\theta^2+\co\left(\theta^3\right)\right\}.}
Clearly, $\ay^{s+r}$ has wandered away from the true valley trajectory
a bit too far, especially for small separations.

We now resume our quest for a better expression for the $I\ib$ valley.
We still want to use linear combinations of the instanton and the
antiinstanton. By now, it should be clear how this can
be done. We put one in the singular gauge and the other in the regular
gauge. We have
\eqn\sprb{A_{I\ib}\ =\ {\rm Im}\left\{
{{\rho_1^2a{\bar x}dx{\bar a}}\over{x^2(x^2+\rho_1^2)}}+
{{b({\overline x-R})dx{\bar b}}\over{(x-R)^2+\rho_2^2}}\right\}.}
Unfortunately, this expression contains some conformal degrees of
freedom, and if we substitute it into \ymaction, these degrees of
freedom do not become zero modes as they should. The brute force
solution to this problem is to use a constraint a la Yung to get rid of
these conformal modes\foot{This constraint gets rid of the zero modes, and
leaves only the quasi-zero modes. This should be compared to the constraints
defined in \yc\ or \lc, which gets rid of both the zero and the
quasi-zero modes, and leaves the
quantum fluctuations.}. We will define $A_{I\ib}$ only on the
slice cut out by this constraint and then conformally project
it to the entire parameter space. For example, if we choose the
constraint \yungslice, we recover $\ay$. There are other obvious
choices of constraints, however. For example, we can use
\eqn\ourslice{\rho_1\ =\ \rho_2\ =\ 1.}
This gives
\eqn\sprc{A_{I\ib}\ =\ {\rm Im}\left\{
{{a{\bar x}dx{\bar a}}\over{x^2(x^2+1)}}+
{{b({\overline {x-R}})dx{\bar b}}\over{(x-R)^2+1}}\right\}.}
To see if this is compatible with the $Z_2$ symmetry
$\theta\rightarrow-\theta$, let's first note that
\eqn\dthetaa{\eqalign{\theta\ &=\ \rho_2-\rho_1\ \ \ \ \ \ \ \ \
{\rm under\ \yungslice,}\cr
&=\ \sqrt{2(z-1)}\ \ \ \ \ {\rm in\ general},\cr
&=\ R\ \ \ \ \ \ \ \ \ \ \ \ \ \ \ \ {\rm under\ \ourslice}.\cr}}
Therefore $\theta\rightarrow-\theta$ is equivalent to $R\rightarrow-R$, which
correponds to moving the $\ib$ from $R$ to $-R$. This can also be achieved
by a rotation, which is a perfectly good symmetry of the expression. Thus we
expect that \sprc\ should respect the $Z_2$ symmetry in question.
Explicit calculation confirms this
expectation. For the \iib\ with opposite phases,
\ie\ $a{\bar b}+b{\bar a}=0$, the action has the small $\theta$ expansion
\eqn\iibactione{S_{I\ib}^{+-}\ \sim\ {{16\pi^2}\over{g^2}}\left\{
2-{1\over3}\theta^2+\co\left(\theta^4\right)\right\}.}
If the phases are aligned with each other,\ie\ $a=b$, we have
\eqn\iibactionc{S_{I\ib}\ \sim\ {{16\pi^2}\over{g^2}}\left\{
{6\over5}\theta^2-{{33}\over{35}}\theta^4+\co\left(\theta^6\right)\right\}.}
Therefore \sprc\ is clearly a better solution than $\ay$.

As mentioned before, the valley solution
has a dependence on the constraint function
$f_z(x)$. It is therefore perfectly plausible for one to discover other
equally satisfactory solutions with
different constraints. One may ask if there
is any reason why he should go through
the trouble of looking for such alternative
solutions. The answer is yes because
eq.\sprc\ in fact gives a divergent field strength
(and consequently a divergent Lagrangian
density) near the origin, even though
the action is a finite and well-behaved
function of $R$. Satisfactory $I\ib$ solutions
with finite field strength everywhere are
not hard to find. For example, we can
choose\foot{We will ignore the
phases $a,b$ again. It is trivial to put
the relative phase back at the end of our
discussion if one chooses to.}
\eqn\sprx{A_{I\ib}\ =\ {\rm Im}\left\{
{{{{\bar x}dx}\over{x^4}}+{{({\overline {x-R}})dx}\over{(x-R)^2}}
\over{1+{{R^2}\over{x^2(1+R^2)}}+{1\over{(x-R)^2}}}}\right\}.}
We will continue to use eq.\sprc, however, not only because we consider
the pathology a mild one, but also because it is easier to generalize
it to $I^N\ib^\nb$ solutions. For those
who are truly bothered by the divergence
problem, eq.\sprc\ and other formulas based on it in this papers could be
viewed as a short-hand for better (but
usually more complicated) solutions such
as eq.\sprx.

\newsec{Multiple Instantons and Antiinstantons}
After dealing with $I\ib$, the generalization to $I^N\ib^\nb$ is relatively
straightforward. Again, we
begin by setting up criteria.
We find that they should read
\vskip 0.2in
\item{1)} $A_{I^N\ib^\nb}$ belongs in the $Q=N-\nb$ sector.
\item{2)} $A_{I^N\ib^\nb}$ has easily identifiable instanton
parameters, and covers the entire $8(N+\nb)$-dimensional
parameter space spanned by all zero- and nonzero-modes.
\item{3.1)} If a subset $I^{N^\prime}\ib^{\nb^\prime}$ becomes
widely separated from the rest, $A_{I^N\ib^\nb}$ reduces to the sum of
$A_{I^{N^\prime}\ib^{\nb^\prime}}$ and
$A_{I^{(N-N^\prime)}\ib^{(\nb-\nb^\prime)}}$.
\item{3.2)} If subsets $I^{N^\prime}$ and
$\ib^{N^\prime}$ have identical sizes and positions, and are widely separated
from the rest, they annihilate each other.
\item{4)} $A_{I^N\ib^\nb}$ respects the conformal symmetries.
\vskip 0.2in
This is rather straightforward once it is written down. The only thing that
needs explanation is that we don't require $I^{N^\prime}$ and
$\ib^{N^\prime}$ to annihilate each other in the presence of other
(anti)instantons. The reason is of course that in the non-trivial background
field generated by other instantons, the parity between instantons and
antiinstantons is broken. This is a manifestation of the nonlinear nature
of the Yang-Mills theory.

Notice that because of $Cri.3.1$, $Cri.3.2$ is equivalent to
\vskip 0.2in
\itemitem{$3.2^\prime$)} If $I^N$ and
$\ib^N$ have identical sizes and positions, $A_{I^N\ib^\nb}$ approaches
the trivial vacuum.
\vskip 0.2in

We will ignore the phases for now. Recall that the $I^N$ solution with
no phases can be written in the 't Hooft form\nref\thoofta{G.'t Hooft,
unpublished.}\nref\corrigan{E.Corrigan and D.B.Fairlie, {\sl Phys. Lett.}
{\bf B67} (1977) 69.}\nref\jackiw{R.Jackiw, C.Nohl and C.Rebbi,
{\sl Phys. Rev.} {\bf D15} (1977) 1642.}\refs{\thoofta{--}\jackiw},
\eqn\ain{A_{I^N}^{\rm 't Hooft}\ =\ {\rm Im}\left\{
{{\sum_{i=1}^N{{\rho_i^2({\overline {x-R_i}})}\over{(x-R_i)^4}}dx}\over
{1+\sum_{i=1}^N{{\rho_i^2}\over{(x-R_i)^2}}}}\right\}.}
This will be the analog of an instanton in the singular gauge.
The analog of an antiinstanton in the regular gauge can be found
by operating on an $\ib^\nb$ solution in the 't Hooft form the following
gauge transformation,
\eqn\gt{g_0\ =\ {{\sum_{i=1}^\nb{{{\rho_i^\prime}^2(x-R_i^\prime)}
\over{(x-R_i^\prime)^4}}}
\over{\Big|\sum_{i=1}^\nb{{{\rho_i^\prime}^2(x-R_i^\prime)}
\over{(x-R_i^\prime)^4}}\Big|}},}
where the ``$\prime$" designates the parameters of the antiinstantons as
compared to those of the instantons. We have
\eqn\aibnb{\eqalign{A_{\ib^\nb}^{g_0}\ =\ &g_0^{-1}
A_{\ib^\nb}^{\rm 't Hooft}g_0+g_0^{-1}dg_0\cr
=\ &{\rm Im}\left\{-\left(
{{\sum_{i=1}^\nb{{{\rho_i^\prime}^2({\overline {x-R_i^\prime}})}
\over{(x-R_i^\prime)^4}}dx}\over
{1+\sum_{i=1}^N{{{\rho_i^\prime}^2}\over{(x-R_i^\prime)^2}}}}\right)\right.\cr
&\left.+{{\left(\sum_{i=1}^\nb{{{\rho_i^\prime}^2({\overline {x-R_i^\prime}})}
\over{(x-R_i^\prime)^4}}\right)
d\left(\sum_{i=1}^\nb{{{\rho_i^\prime}^2(x-R_i^\prime)}
\over{(x-R_i^\prime)^4}}\right)}
\over{\Big|\sum_{i=1}^\nb{{{\rho_i^\prime}^2(x-R_i^\prime)}
\over{(x-R_i^\prime)^4}}\Big|^2}}\right\}.\cr}}
Now, clearly the first term in \aibnb\ exactly cancels \ain\ when
the positions and sizes of the instantons are identical to those of the
antiinstantons. Thus if we choose
\eqn\ainibnb{A_{I^N\ib^\nb}\ =\ A_{I^N}^{\rm 't Hooft}+A_{\ib^\nb}^{g_\nb},}
it will satisfy $Cri.3.2^\prime$. In fact, it is easy
to see that it also satisfies $Cri.3.1$ because if some
(anti)instantons are far away, their contributions are suppressed
by at least the inverse square of the distances, in both the numerator and
the denominator of the expression.

As for the other criteria, $Cri.1$ and $2$ are again satisfied trivially.
$Cri.4$ requires more
thought, though. Clearly \ainibnb\ respects the translational and
rotational symmetries. The special conformal transformations will introduce
relative phases within any pair in either of the subsets $I^N$ or
$\ib^\nb$ unless the vector of the special conformal boost coincides with
the axis of the $I^2$ ($\ib^2$) pair\wanga. Since we have assumed no relative
phase so far, we don't have to worry about these special conformal
transformations except for a few special cases, such as $I^2\ib$ or when
everything lines up in a straight line. In either case, one simply introduces
any appropriate constraint to kill off the extra degree of freedom. The
same can be easily done for dilitation also.
Anyway, we can be excused for skimping the details concerning the dilitation
and the special
conformal symmetries because they are not present in the \ymh\ theory
wherein our ultimate interest lies.

With \ainibnb, one may begin by computing $S(A_{I\ib})$. Subtracting
the ``self-action" $2S_I$ from $S(A_{I\ib})$ then gives the two-body
interaction between an \iib\foot{Note that because we have used the exact
$N$-instanton solution in our construction, the interaction among any number
of instantons remains zero. The same is true for antiinstantons.}.
One then proceeds to compute $S(A_{I^2\ib})$ and
$S(A_{I\ib^2})$. Subtracting the self-action
and the two-body interactions between
all pairs then gives the three-body interactions. This process can be carried
over to yield the $n$-body interaction for any $n$. In practice, one may want
to assume that these many body interactions become less and less important as
$n$ grows large.

We have given the expressions for the $I^N\ib^\nb$ valley configurations
without phases. Now we will see how to introduce phases into them.
The two overall phases $a$ and $b$ for $I^N$ and $\ib^\nb$ respectively
can be put into \ainibnb\ in the same manner as in \spra. The relative
phases within $I^N$ ($\ib^\nb$) are much harder to deal with, however.
As readers familiar with ref.\xref\atiyah\ would
know, the $8N-3$ physical parameters
of the exact $I^N$ solution are buried deep in a maze of quaternion
matrix algebra. To interpret the positions, sizes and phases of even the
simplest $I^2$ solution is not exactly a trivial task\wanga.
It is therefore not surprising to find that our linear construction
of the $I^N\ib^\nb$ valley doesn't work with these solutions.
More specifically, we are unable to find the suitable gauge transformation
as in \gt\ which
is vital for our solution to satisfy $Cri.3.2^\prime$.

Although this looks very much like the end of the story, we in fact have
another recourse to go to. This is the work of Jackiw, Nohl and
Rebbi\jackiw, in which they generalized the 't Hooft form to include
more parameters, \ie
\eqn\ainj{A_{I^N}^{\rm JNR}\ =\ {\rm Im}\left\{
{{\sum_{i=0}^N{{\lambda_i^2({\overline {x-r_i}})}\over{(x-r_i)^4}}dx}\over
{\sum_{i=0}^N{{\lambda_i^2}\over{(x-r_i)^2}}}}\right\}.}
We shall call this the JNR gauge because it is gauge-equivalent to
other forms of the $I^N$ solution. Notice that the overall
scale of $\lambda$'s gets canceled between the numerator and the denominator,
so there seems to be a total of $5N+4$ parameters now.
More careful examination reveals that some of these parameters correspond
to gauge degrees of freedom for $N\le2$, so the actual numbers of
independent parameters are 5 and 13 for $N=1$ and 2 respectively.

Although it is not obvious from looking at \ainj, the extra parameters
it carries compared to the 't Hooft form in fact correspond to relative
phases\wanga. Amazingly, \ainj\ doesn't contain any quaternion matrices,
and the analog of $g_0$ as in \gt\ can indeed be found. A
discussion similar to what we did with the 't Hooft form then follows.
We again skimp the details for the following reasons. The algebra
is very messy and not inspiring at all. The problem it
solves is not particularly important either, since
when we evaluate a path integral, the integral over the phases can
usually be approximated with the group volume.
Besides, for large $N$'s, \ainj\ clearly doesn't have enough parameters
to cover all the phases. We therefore simply state without proving the
following result. Satisfactory expressions for $I^2\ib^2$ and $I^3\ib^3$
covering the entire parameter space can be found using \ainj. It may
seem strange at first that it would work for $I^3\ib^3$, since
the JNR form \ainj\ is 2 parameters short for the entire space of $I^3$.
Fortunately the conformal degrees of freedom are more than enough to
make up for the difference.

\bigbreak\bigskip\bigskip\centerline{{\bf Acknowledgements}}\nobreak
We thank S.Weinberg for pointing out the
necessity to include $Cri.1$ in the
criteria in both Section 3 and Section 4. This research is supported by
the theory group at University of Texas at Austin.

\appendix{A}{Quaternions}
Similar to its $\cc$-number
cousin $z=z_0+iz_1$, a quaternion $x\in \calh$ and
its conjugate $\xb$ are given by
\eqna\dq
$$\eqalignno{x&=x_0+ix_1+jx_2+kx_3,&\dq a\cr
\xb&=x_0-ix_1-jx_2-kx_3,&\dq b\cr}$$
where $x_\mu\in \calr$, and $\{i,j,k\}$ satisfy
\eqn\dijk{\eqalign{&i^2=j^2=k^2=-1,\cr
&ij=-ji=k,\ \ \ jk=-kj=i,\ \ \ ki=-ik=j.\cr}}
Clearly the quaternion algebra
has a $2\times2$ complex matrix representation :
\eqn\iijk{\{1,i,j,k\}\ \rightarrow\ \{I,i{\vec \sigma}\},}
where $\sigma_m$ are the Pauli matrices. Therefore the group $SU(2)$ can be
identified with $SP(1)$, the group
of unit quaternions, and the $SU(2)$ algebra
correspond to ${\rm Im}\ \calh$.

One can also identify $\calr^4$ with $\calh$ via \dq{a}, and the $SU(2)$
gauge field $A_\mu(x)$ is then obviously a function of quaternions with
imaginary quaternion values. When working with Yang-Mills instantons,
we find that the notation can be even further simplified
if we consider the one-form
\eqn\gof{A(x)\ =\ \sum_{\mu=0}^3{A_\mu(x)dx^\mu}.}
The BPST instanton traditionally expressed in terms of the 't Hooft $\eta$
tensor as
\eqn\etains{A_\mu(x)\ =\ \sum_{\mu=0}^3{{\sigma^m\eta_{m\mu\nu}x^\nu}
\over{i(x^2+\rho^2)}},}
can now be written as
\eqn\quains{A_I(x)\ =\ {\rm Im}\left\{{{xd\xb}\over{x^2+\rho^2}}\right\},}
and the antiinstanton is
\eqn\quaains{A_\ib(x)\ =\ {\rm Im}\left\{{{\xb dx}\over{x^2+\rho^2}}\right\}.}

It is possible to do computations in the quaternion notation. For example,
one may wish to evaluate the curvature 2--form $F$ for the gauge field defined
in \quains. It is given by
\eqn\fstr{\eqalign{F\ &=\ dA+A\wedge A\cr
&=\ {\rm Im}\left\{\left({{dx\wedge d\xb}\over{x^2+\rho^2}}
-{{xd(x^2+\rho^2)\wedge d\xb}\over{(x^2+\rho^2)^2}}\right)+
{{xd\xb\wedge xd\xb}\over{(x^2+\rho^2)^2}}\right\}\cr
&=\ {\rm Im}\left\{{{dx\wedge d\xb}\over{x^2+\rho^2}}
-{{x(d\xb x+\xb dx)\wedge d\xb}\over{(x^2+\rho^2)^2}}+
{{xd\xb\wedge xd\xb}\over{(x^2+\rho^2)^2}}\right\}\cr
&=\ {{\rho^2dx\wedge d\xb}\over{(x^2+\rho^2)^2}}.\cr}}
We dropped the ${\rm Im}$ symbol in
the final expression because it is already
pure imaginary.

A slightly more complicated example
is to examine how \quains\ transforms under
a special conformal boost, which can be defined as
\eqn\scb{x\ \rightarrow\ \xp\ =\ (x+a)(1-\ab x)^{-1}.}
We begin by inversing \scb,
\eqn\scba{x\ =\ (1+\xp\ab)^{-1}(\xp-a)\ =\ (\xp-a)(1+\ab\xp)^{-1}.}
Substituting \scba\ into \quains, one finds that
\eqn\quainsa{A_I(x)\ =\ {\rm Im}\left\{{{(1+\xp\ab)^{-1}(\xp-a)
d\left[(\xpb-\ab)(1+\xp\ab)\right]}
\over{(\xp-a)^2+\rho^2(1+\xp\ab)^2}}\right\}.}
This can be simplified with a gauge transformation,
\eqn\gta{g\ =\ {{1+a\xpb}\over{\big|1+a\xpb\big|}}.}
We have
\eqn\quainsb{\eqalign{A\rightarrow A^\prime=&g^{-1}Ag+g^{-1}dg\cr
=&{\rm Im}\left\{{{(\xp-a)d\left[(\xpb-\ab)(1+\xp\ab)\right](1+\xp\ab)^{-1}}
\over{(\xp-a)^2+\rho^2(1+\xp\ab)^2}}+
{{(1+\xp\ab)ad\xpb}\over{(1+a\xpb)^2}}\right\}\cr
=&{\rm Im}\left\{{{(\xp-a)d\xpb}\over{(\xp-a)^2+\rho^2(1+\xp\ab)^2}}\right.
-{{(\xp-a)^2(1+\xp\ab)ad\xpb}\over
{(1+a\xpb)^2[(\xp-a)^2+\rho^2(1+\xp\ab)^2]}}\cr
&\left.\ \ \ \ \ +{{(1+\xp\ab)ad\xpb}\over{(1+a\xpb)^2}}\right\}\cr
=&{\rm Im}\left\{{{(\xp-a)d\xpb}\over{(\xp-a)^2+\rho^2(1+\xp\ab)^2}}
+{{\rho^2(1+\xp\ab)ad\xpb}\over{(\xp-a)^2+\rho^2(1+\xp\ab)^2}}\right\}\cr
=&{\rm Im}\left\{{{(\xp-R)d\xpb}\over
{(\xp-R)^2+{\rho^\prime}^2}}\right\},\cr}}
where
\eqn\drr{R\ =\ {{(1-\rho^2)a}\over{1+\rho^2a^2}}\ \ \ \ \ {\rm and}
\ \ \ \ \ \rho^\prime\ =\ {{(1+a^2)\rho}\over{1+\rho^2a^2}}.}
This gives how the parameters of a single instanton
change under the special conformal
transformation.
\listrefs
\bye